\documentclass[12pt]{article}

\usepackage{jheppub}
\usepackage{amsfonts,amsmath} 
\usepackage{mathrsfs} 
\usepackage{setspace}
\usepackage[utf8]{inputenc} 
\usepackage{comment} 

\makeatletter
\def\@fpheader{\relax}
\makeatother

\newcommand{\be}{\begin{equation}} \newcommand{\ee}{\end{equation}}

\newcommand{\Dslash}{\,\ensuremath \raisebox{0.7mm}{\slash}
\hspace{-2.8mm}\nabla\hspace{0.2mm}}

\title{On the Six-dimensional Kerr Theorem and Twistor Equation}

\author{Bruno Carneiro da Cunha}
\emailAdd{bcunha@df.ufpe.br}
\affiliation{Departamento de Física, Universidade Federal de Pernambuco,
50670-901, Recife, Pernambuco, Brazil} 

\abstract{The Kerr theorem is revisited as part of the twistor program
  in six dimensions. The relationship between pure spinors and
  integrable 3-planes is investigated. The real condition for
  Lorentzian spacetimes is seen to induce a projective property in the
  space of solutions, reminiscent of the quaternionic structure of the
  6-dimensional Lorentz group. The twistor equation (or Killing spinor
  equations generically) also has an interpretation as integrable null
  planes and a family of Einstein spacetimes with this property are
  presented in the Kerr-Schild fashion.} 

\preprint{\today}

\begin{document}

\maketitle

\section{Introduction}

The Kerr theorem shows how to generate maximally null integrable
submanifolds in flat spacetime. The importance of such
characterization was technical at first, with solutions of Einstein's
Equations in the Newman-Penrose formalism in mind
\cite{Penrose1985,Penrose1986}. Many solutions have been found this
way, including the Kerr-Newman solution. For complexified spaces,
Hughston and Mason \cite{Hughston:1988nz} showed how to parametrize
these null manifolds in terms of pure spinors, and posteriorly related
the structure to the integrability of the wave equation
\cite{Hughston:1990-1,Hughston:1990-2}. The same connection was noted
by the mathematicians at around the time \cite{Kurusa:1991} to provide
a higher-dimensional analogue of ``Bateman's Formula'', or Penrose
transform. These have been recast in the twistor language by Berkovits
and Cherkis \cite{Berkovits:2004bw}. 

From the physical point of view, the existence of null integrable
submanifolds is at the heart of the simplification of the metric
structure that happens near the horizon of a black hole, even without
the presence of supersymmetry,
\cite{Chamseddine:1996pi,daCunha:2013mha}. This is a horizon analogue
of the ``peeling theorem'' \cite{Wald}, which originally stated that the
corrections to the metric in asymptotic flat spacetimes are ranked in
order of algebraic specialty, as defined by Petrov. The Petrov
classification of the Weyl tensor sees the latter as a linear
application in the space of two vectors. Because of the symmetry of the
Weyl tensor, eigenvalues are the outer product of null vectors, dubbed
principal null directions. With respect to the peeling theorem,
in particular, one can state that the contributions to the metric from
charges such as ADM mass and angular momentum allow for the definition of
a principal null direction, picking a particular one from the infinite
directions of flat spacetime. This fact was used by Penrose to define
local conserved charges in Twistor Theory \cite{HuggettTod,WardWells}. 

The relationship between null integrable manifolds and spinors have
been outlined since its inception \cite{Cartan:1981,Chevalley:1954}. In
the applications of spinors in General Relativity this connection has
always been explicit, however, applications in supergravity and
superstrings the algebraic approach has always been preferred. In four
dimensions, the work of van Niuwenhuizen and Warner
\cite{vanNieuwenhuizen:1983wu} built the bridge between the two formalisms
by working the conditions for the existence of Killing spinors in the
Newman-Penrose formalism. 

This letter will try a similar program in six dimensions. We will
revisit the Kerr theorem in six dimensions and define helicity spinors
by exploring the symmetry of the solutions. Previous work along those
lines was conducted in \cite{Cheung:2009dc}. We introduce the twistor
equation in 6 dimensions as a sufficient condition for the existence
of integrable isotropic 3-planes. We close with the relation to the
existence of Killing spinors and applications in six-dimensional
spaces to a family of examples presented in the Kerr-Schild form. 

\section{Six Dimensional Newman-Penrose Formalism}

We begin by reviewing some aspects of four dimensional spinors. There,
the spinor formalism takes full advantage from the fact that the
Lorentz group $SO(3,1)$ is isomorphic to $SU(2)\times SU(2)$. Reality
conditions relate both factors, so the decomposition of a vector into
$SU(2)$ pieces is done via the van der Waerden symbols to a matricial
object $A_{\alpha\dot{\alpha}}$. The norm of the vector is written as
the determinant of the matrix. When it is zero, it means that both
rows are linearly dependent, so that:  
\be
A^{\alpha\dot{\alpha}}=\kappa^\alpha\bar{\kappa}^{\dot{\alpha}}.
\ee 
the objects $\kappa^\alpha$ and $\bar{\kappa}^{\dot{\alpha}}$ transform as
spinors under Lorentz, and parametrize the null vectors. $\kappa$ and
$\bar{\kappa}$ will be the complex conjugate of each other if we want
the null vector to be real. The correspondence between spinors and null
vectors is one up to a multiplicative phase, the ``flag'' of the spinor,
which in itself can be used to encode helicity information, a fact which
was much useful for the computation of solutions of arbitrary spin
massless fields and their scattering amplitudes. 

One can go ahead and substitute the tetrad formalism with spinor language
in general relativity. Let $e_0,e_1,e_2,e_3$ be a normalized tetrad. The
null vectors 
\be
k=(e_0+e_1)/\sqrt{2},\quad l=(e_0-e_1)/\sqrt{2},\quad
m=(e_2+ie_3)/\sqrt{2},\quad \bar{m}=(e_2-ie_3)/\sqrt{2}
\ee
can be decomposed with two (complex) spinors: $\imath^\alpha$ and
$o^\alpha$ and their complex conjugate:
\be
k^{\alpha\dot{\alpha}}=\imath^{\alpha}\bar{\imath}^{\dot{\alpha}},\quad
l^{\alpha\dot{\alpha}}=o^\alpha\bar{o}^{\dot{\alpha}},\quad
m=\imath^{\alpha}\bar{o}^{\dot{\alpha}},\quad
\bar{m}=o^{\alpha}\bar{\imath}^{\dot{\alpha}}.
\ee
Note that $k^a$ and $l^a$ are real whereas $m^a$ is the complex conjugate
of $\bar{m}^a$.  

In six dimensions the decomposition of vectors is similar
\cite{Batista:2012cp}. The symmetry
group is $SO(6)$, which we will see as a complex group $SL(4,\mathbb{C})$.
One real form of it is the group of isometries of Euclidian space,
$SO(6)$, the compact form of
$SL(4,\mathbb{C})$ and is isomorphic locally to
$SU(4)$\footnote{\footnotesize Globally $SU(4)$ a double cover of
$SO(6)$.}. The latter has a natural 
 4-dimensional fundamental ${\bf 
4}$ representation which can be seen as the spinorial representation of
$SO(6)$. Paralleling the 4-dimensional case, a six dimensional vectorial
representation can be constructed from the tensor product of the spinorial
representation: 
\be
{\bf 4}\times {\bf 4} = {\bf 6} + {\bf 10}
\ee
and then a vector can be written as an antisymmetric tensor of spinor
indices:
\be
V^{AB}=\Sigma^{AB}_aV^a
\ee
where $\Sigma^{AB}_a=-\Sigma^{BA}_a$ are analogues of the van der Waerden
symbols in six dimensions, essentially the positive chirality projection
of the Dirac matrices. The inner product of vectors is performed using
the completely antisymmetric symbol $\epsilon_{ABCD}$: 
\be
V^aV_a\equiv \epsilon_{ABCD}V^{AB}V^{CD}.\label{eq:dotprod}
\ee
Now, let us consider a spinorial basis $\kappa_i^A$, normalized so that
$\epsilon_{ABCD}\kappa_1^A\kappa_2^B\kappa_3^C\kappa_4^D=1$. A generic
vector can be expanded in a basis of simple bispinors 
$\kappa^{[A}_i\kappa^{B]}_j$, but, because the $\epsilon_{ABCD}$ is
completely antisymmetric, the only chance that this vector will not be
null is that the expansion involves all of the $\kappa_i$. One can then
find, for a given null vector $V^{[AB]}$, a spinor basis
$({\kappa_i})^A$, $i=1,2,3,4$ such that $V^{[AB]}$ can be decomposed in a
bispinor basis involving only three of the $(\kappa_i)^A$:
\be
V^{AB}=\alpha_1\kappa_1\wedge \kappa_2+\alpha_2
\kappa_1\wedge\kappa_3 + \alpha_3 \kappa_2\wedge\kappa_3
=(\kappa_1+\frac{\alpha_3}{\alpha_2}\kappa_2)\wedge
(\alpha_1\kappa_2+\alpha_2\kappa_3)=\chi^{[A}\xi^{B]},
\ee
supposing $\alpha_2\neq 0$. Then,
paralellizing the result in 
four dimensions, any null vector can be represented as the outer product
of spinors, with the converse obviously true, given \eqref{eq:dotprod}.

While a vector is associated with the exterior product of two spinors, a
single spinor $\kappa^A$ is associated with an isotropic 3-plane. This
3-plane is
generated by the null vectors
\be
V_i^{AB}=\kappa^{[A}\kappa_i^{B]},
\ee
for some basis $\kappa_i^A$. It is clear that there are 3 nonzero vectors
$V_i^{AB}$ and that they generate an isotropic space, since any linear
combination of the $V_i^a$ is a null vector. 

A tangent space connection $\nabla_a$ can be introduced, and from the
zero torsion condition and the Leibniz rule it follows that the following
operator is linear on spinorial fields:  
\be
(\nabla_a\nabla_b-\nabla_b\nabla_a)\kappa^E={{\cal R}_{abF}}^E\kappa^F.
\ee 
where $a$ and $b$ are coordinate (vector) indices. Since it is
antisymmetric under interchange between $a\equiv AE$ and $b\equiv FG$,
one can construct from it the spinorial tensor:
\be
{R_{AB}}^{CD}=\epsilon^{EFGC}{{\cal R}_{AEFGB}}^{D}
\ee
which is the spinor analogue of the Riemann tensor. As in four
dimensions, one can decompose it into isometry invariant pieces: 
\be
{R_{AB}}^{CD}=\Lambda
 (\delta_A^C\delta_B^D-4\delta_B^C\delta_A^D)+ {\Phi_{AB}}^{CD}
+{\Psi_{AB}}^{CD}\label{eq:riemanndecomp}
\ee
with $\Lambda$ related to the scalar
curvature, ${\Phi_{AB}}^{CD}={\Phi_{[AB]}}^{[CD]}$ with
${\Phi_{AC}}^{AD}=0$ being the spinor analogue of the traceless part
of the Ricci tensor and ${\Psi_{AB}}^{CD}={\Psi_{(AB)}}^{(CD)}$ with
${\Psi_{AC}}^{AD}=0$ being the spinor analogue of the Weyl tensor. The
odd form of the first term guarantees that ${R_{AB}}^{AD}=0$, while
${R_{AB}}^{CA}=-15\Lambda\delta_B^C$. In terms of $SU(4)$ irreducible
representations, the decomposition is ${\bf 1} + {\bf 20} + {\bf 84}$,
adding up to $105$ independent components of the Riemann tensor in six
dimensions. 

\section{\label{se:kerr}The Kerr Theorem}

The Kerr theorem gives an implicit solution to all analytic,
shear-free, geodetic null congruences in flat space. The problem is
strongly tied with the existence of integrable null submanifolds. The
connection between the two seemingly distinct problems is deep, as will
be discussed in the subsection on generic dimensions. The original result,
in four dimensions, can be written in terms of an implicit, analytic
function in projective twistor space, and actually serves as one
geometrical interpretation of a twistor \cite{HuggettTod}. We will see
that most of the geometrical picture translate to six dimensions. 

We will follow the discussion in \cite{Stephani2003b}, taking a
vectorial point of view. This problem has also been considered using 
spinorial techniques for the general even-dimensional case by 
\cite{Hughston:1988nz}. For a more recent consideration see
\cite{TaghaviChabert:2012uj}. Let us begin with flat 6-dimensional
flat space with Lorentzian signature: 
\be
ds^2=dudv+dz_1d\bar{z}_1+dz_2d\bar{z}_2.
\ee
The null 1-forms:
\be
e_0=du+\bar{Y}^idz_i+Y^id\bar{z}_i-Y_i\bar{Y}^idv,
\quad e_i=dz_i-\bar{Y}_idv
\ee
will determine an integrable distribution (submanifold) if their
brackets all close, or, in the dual formulation of Frobenius' theorem, 
if $de^i$ can be written as a linear combination of $e^i\wedge e^j$
tangent to the submanifold. A somewhat tedious computation yields, for
$de_0\wedge e_0 \wedge e_1 \wedge e_2=0$ the following equations:  
\be
\begin{gathered}
\partial_v Y^1 + Y^2\partial_2Y^1+Y^1\partial_1Y^1-Y^1\bar{Y}^1
\partial_u Y^1-
Y^1\bar{Y}^2\partial_uY^2+\bar{Y}^1\bar{\partial}_1Y^1+\bar{Y}^2
\bar{\partial}_1
Y^2 = 0, \\
\partial_v Y^2 + Y^2\partial_2Y^2+Y^1\partial_1Y^2-Y_2\bar{Y}^1 \partial_u
Y^1-
Y^2\bar{Y}^2\partial_uY^2+\bar{Y}^1\bar{\partial}_2Y^1+\bar{Y}^2
\bar{\partial}_2
Y^2 = 0, \\
\bar{\partial}_2Y^1-\bar{\partial}
_1Y^2+Y^1\partial_uY^2-Y^2\partial_uY^1=0,\\ 
Y^2(\partial_vY^1+\bar{Y}^2\bar{\partial}_1Y^2-Y^1\partial_2Y^2+Y^2
\partial_2Y^1
+\bar{Y}^1\bar{\partial}_1Y^1) \quad\quad\quad\quad\quad\quad\quad\quad
\\ 
\quad\quad\quad\quad\quad\quad\quad\quad -Y^1(\partial_vY^2
+\bar{Y}^1\bar{\partial}_2Y^1+Y^1\partial_1Y^2-Y^2\partial_1Y^1+
\bar{Y}^2\bar{\partial}_2
Y^2)=0.
\end{gathered}
\ee
and the equations following from the requisition that $de_1\wedge e_0
\wedge e_1 \wedge e_2$ and $de_2\wedge e_0 \wedge e_1 \wedge e_2$
vanish:
\be
\begin{gathered}
\bar{\partial}_1Y^i-Y^1\partial_uY^i=0,\quad\quad
\bar{\partial}_2Y^i-Y^2\partial_u Y^i=0,\\
Y^2\bar{\partial}_1Y^i-Y^1\bar{\partial}_2Y^i=0,
\end{gathered}
\ee
with $i=1,2$. Note that the first two equations imply the third one.

Despite the coupling, the equations are simple to solve. After some
manipulation, they result in
\be
(\bar{\partial}_i-Y^i\partial_u)Y^j=0,\quad\quad
(\partial_v+\sum_iY^i\partial_i)Y^j=0,\quad i=1,2,
\label{eq:int6d}
\ee
and their complex conjugate. These equations can be solved by the
method of characteristics \cite{John1991}. The solution is given
implicitly in terms of a general function $F$ of five complex
arguments: 
\be
F\left(Y^1,Y^2;vY^1-z_1,vY^2-z_2;u+\bar{z}_1Y^1+\bar{z}_2Y^2\right)=0.
\label{eq:solutionkerr}
\ee
This version of the solution of integrable distributions is more
economical than the general expression given in
\cite{Hughston:1988nz}, but the spinorial interpretation is less
clear. In order to recover it, let us introduce the spinors
\be
\kappa^A=
\left[
\begin{array}{c}
Y^1 \\ Y^2 \\ 0 \\ 1 
\end{array}
\right],
\quad\quad
\bar{\kappa}^A=
\left[
\begin{array}{c}
\bar{Y}^2 \\ -\bar{Y}^1 \\ 1 \\ 0 
\end{array}
\right].
\ee
which generate the null vector $k^a\equiv k^{[AB]}$ by exterior
product: $k^{AB}=\kappa^{[A}\bar{\kappa}^{B]}$. Note that
$\bar{\kappa}^A$ is related to the complex conjugate of $\kappa^A$ via
a conjugation operator:
\be
\bar{\kappa}^A=B^{AB}\kappa^*_B,\quad\quad B =
\left[
\begin{array}{cc}
i\sigma^2 & 0 \\
0 & i\sigma^2 
\end{array}
\right]
= 1\otimes i\sigma^2,
\label{eq:brep}
\ee
so that the vector $k^a$ is real. 

Lastly, we introduce the ``position
vector'' $x_{AB}$, an antisymmetric matrix in spinor space:
\be
x_{AB}=
\left[
\begin{array}{cccc}
0 & v & -\bar{z}_1 & -z_2 \\
-v & 0 & -\bar{z}_2 & z_1 \\
\bar{z}_1 & \bar{z}_2 & 0 & u \\
z_2 & -z_1 & -u & 0
\end{array}
\right].
\ee
The attribution of the positions of the coordinates is of course
arbitrary. There is a $SU(2)\times SU(2)\simeq SO(4)$ isometry -- the 
``little group'' -- in spinor space which keeps the vector
invariant. We will discuss its geometrical interpretation. Recasting
\eqref{eq:solutionkerr} using spinors: 
\be
F(\kappa^A,x_{CB}\kappa^B)=0,
\label{eq:projtwistors}
\ee
where $F(\kappa^A,\zeta_B)$ is a holomorphic function from ${\bf
  4}\times {\bf \bar{4}}$ to the complex numbers.  The pair
$Z^{{\cal
    I}}=(\kappa^A,\zeta_B)$ is a six-dimensional twistor. From the
construction above, the function $F$ needs only be defined for
$\kappa^A\zeta_A=0$, so one defines in twistor space a natural
pairing: 
\be
\langle Z^{\cal I},U^{\cal J}\rangle \equiv \langle (\lambda^A,\chi_B),
(\pi^C,\zeta_D) \rangle =\lambda^A\zeta_A+\pi^D\chi_B.
\ee
With the choice of conjugation $\bar{Z}^{\cal
  I}=(\bar{\kappa}^A,\bar{\zeta}_B)$,
$\bar{\zeta}_B=(B^{-1})_{BC}(\zeta^*)^C$, this form has signature
$(6,2)$.  

The crucial fact in six dimensions is that now $F$ has the projective
property: 
\be
F(\alpha\kappa^A+\beta\bar{\zeta}^A,\alpha\zeta_B+\beta\bar{
\kappa}_B)=F(\kappa^A , \zeta_B).
\ee
These transformations keep the vector
$k^{AB}=\kappa^{[A}\bar{\kappa}^{B]}$ invariant. This property shows
that the relation between twistors and geodesic, shear-free null
vector fields is not bijective, as in \cite{Hughston:1988nz}. There
are many isotropic 3-planes which give rise to the same integrable
vector field. This contrasts to the situation in $3+1$ dimensions,
where, given a real null vector field, all isotropic planes containing
it are related by a phase. It should also be pointed out that without
the reality condition enforced by the operator $B$, the solution $F$
has to be thought of as a projective section on the Grassmanian space
$Gr(2,4)$, which, unlike its 4-dimensional sibling, is not a
projective space.  

Incidentally, in the six dimensional case the antisymmetric operator
$x_{AB}$ is not determined uniquely from $\kappa^A$ and $\zeta_A$,
since the spinor space has more than two dimensions. The notion that
the space-time coordinates arise from the ``more fundamental''
objects, the twistors, cannot work as in four dimensions.  

\subsection{\label{se:6dspinors}Six Dimensional Helicity Spinors} 

The results above encode an underlying algebraic structure. The
six-dimensional Lorentzian space considered also shows up in the
construction of instantons for four dimensional Yang-Mills theories,
where the space of solutions can be seen as the twistor space for
Euclidian four dimensional spacetime \cite{Atiyah:1979iu}. The
construction is relevant for our case. Let us recall that the twistor
space in four dimension is defined, {\it e. g.} in \cite{HuggettTod}
as the pair of 2-spinors $Z^a=(\omega^\alpha,\pi_{\dot{\beta}})$ with
a natural linear symmetric inner product of signature $(2,2)$: 
\be
\langle Z_1,Z_2 \rangle = (\omega_2)^\alpha(\bar{\pi}_1)_\alpha +
(\bar{\omega}_1)^{\dot{\alpha}}(\pi_2)_{\dot{\alpha}}.
\ee 
The group that keeps this structure invariant is another real form of 
$SL(4,\mathbb{C})$, $SU(2,2)$, which is isomorphic to $SO(4,2)$. For
$SO(5,1)$, the
accidental isomorphism is with $SL(2,\mathbb{H})$, or $2\times 2$,
unit-determinant matrices over the quaternion field $\mathbb{H}$
\cite{Atiyah:1979iu,Kugo:1982bn}. Thus, spinor space in 6 dimensions
is essentially another real form of twistor space in four
dimensions. Given four complex numbers $\{z_1,z_2,z_3,z_4\}$, one can
embed them into a pair of quaternions as follows: 
\be
(z_1,z_2,z_3,z_4)\rightarrow (z_1+jz_2,z_3+jz_4),
\ee 
with $j\neq i$ a basis element of the quaternions.

In the quaternion language, the symmetry
\be
\left(
\begin{array}{c}
\kappa^A \\
\zeta_B
\end{array}
\right)
\equiv
\left(
\begin{array}{c}
\alpha\kappa^A+\beta\bar{\zeta}^A \\
\alpha\zeta_B+\beta\bar{\kappa}_B
\end{array}
\right)
\ee
corresponds to right multiplication by a quaternion $q=\alpha+\beta j$
\cite{Kugo:1982bn}. Left and right multiplication from unit-norm
quaternions make for a ${\rm SU(2)}\times {\rm SU(2)}\sim {\rm SO(4)}$
symmetry that acts on the spinor $\kappa^A$ but that leaves the null
vector $\kappa^{[A}\bar{\kappa}^{B]}$ invariant. In terms of the
particular spinor representation given in \eqref{eq:brep}, the right
action is of the form $g\otimes i\sigma^2$, $g\in {\rm SU(2)}$, the
generic operator which commutes with $B$. This symmetry is the
``little group'' of $SO(5,1)$, and, among the multiple spinors which
represent the same null vector $k^{AB}$, one line is chosen by
representing helicity, a space-like vector orthogonal to
$k^{[AB]}$. The function representing the solution of the Kerr theorem
is actually a function on $\mathbb{HP}^1$, the projective quaternionic
line \cite{Atiyah:1979iu}. 

To each isotropic and simple 3-plane one associates a spinor: given a
null basis $\{e^1,e^2,e^3\}$ of the 3-plane, one constructs a basis of
spinors $(\kappa_i)^A$ such that $e^i=(\kappa_0)^{[A}(\kappa_i)^{B]}$
for each $i=1,2,3$. Following the treatment in \cite{Batista:2012cp},
given a spinor $\kappa^A$, there is only one isotropic 3-plane
associated to it: the 3-form $T_{abc}\equiv T_{[AB]\,[CD]\,[EF]}$ obtained
by inverting the relation: 
\be
T_{ABCDEF}\epsilon^{DEFG}=\bar{T}_{A[B}\delta_{C]}^G+
\tilde{T}^{FG}\epsilon_{ABCF}  
\ee 
defines a generic 3-plane for symmetric $\bar{T}_{AB}$ and
$\tilde{T}^{AB}$. If one has either of them simple, and the other
zero, like $\bar{T}_{AB}=0$ and $\tilde{T}^{AB}=\kappa^A\kappa^B$, the
3-plane is isotropic (and self-dual). This 3-plane contains the null
vector $k^a=\kappa^{[A}\bar{\kappa}^{B]}$. The vector is actually in the
intersection of the 3-planes associated with $\kappa^A$ and
$\bar{\kappa}^A$.  

In order to define a helicity of a twistor, take the pair
$(\kappa^A,\zeta_B)$ and construct from it the 2-plane:
\be
S^{[AB][CD]}=\kappa^{[A}\epsilon^{B]CDE}\zeta_E-
\kappa^{[C}\epsilon^{D]EBC}\zeta_E. 
\ee
Because of the null condition $\kappa^A\zeta_A=0$, this plane is
simple and  contains the vector
$k^a=\kappa^{[A}\bar{\kappa}^{B]}$. The other vector can be extracted
with the following procedure. Define the rest of the spinor basis
$\pi$ and $\bar{\pi}$ such that 
\be
\kappa^{[A}\bar{\kappa}^B\pi^C\bar{\pi}^{D]}=\epsilon^{ABCD}.
\ee
This essentially means that $\pi^A$ is dual to $\zeta_A$. We have the
helicity as the vector: 
\be
m^a=\kappa^{[A}\pi^{B]}+\bar{\kappa}^{[A}\bar{\pi}^{B]}.
\ee
Likewise, one can associate with the coordinates of $\mathbb{HP}^1$
the null vectors and all helicity vectors associated with the little
group. 

Given this, one can reinterpret formulas given in
\cite{Berkovits:2004bw} as holomorphic projective sections in twistor
space, or $\mathbb{HP}^1$. It would be interesting if one can use
analytic properties to define n-point functions of 6-d Lorentz
invariant massless field theories, as in
\cite{Weinberg:2010fx,Weinberg:2012mz,Saemann:2011nb}. The
construction seems similar to that of \cite{Cheung:2009dc}, but with
the extra geometric interpretation.

\subsection{A Digression in Generic (even) Dimensions}

In a flat, even dimensional Lorentzian manifold of null coordinates
$\{u,v,z^i,\bar{z}^i\}$, the analogue of \eqref{eq:int6d} is: 
\be
(\bar{\partial}_i-Y^i\partial_u)Y^j=0,\quad\quad
(\partial_v+\sum_iY^i\partial_i)Y^j=0,\quad i,j=1,\ldots,(D-2)/2,
\label{eq:intgeneral}
\ee
and their complex conjugate. Given the vector fields
\be
k^a=\frac{\partial}{\partial v}+Y^i\frac{\partial}{\partial
  z^i}+\sum_i\bar{Y}^i(m^i)^a,\quad\quad
(m^i)^a=\frac{\partial}{\partial
  \bar{z}^i}-Y^i\frac{\partial}{\partial u}.
\ee
We can see that \eqref{eq:intgeneral} are equivalent to the vanishing 
of the ``anti-holomorphic'' part of the (flat space covariant)
derivative of $k^a$: 
\be
(m^i)^b(m^j)^a\partial_ak_b=0,\quad\quad i,j=1,\ldots,(D-2)/2.
\ee
which in turn means that the attribution of ``holomorphic'' and
``anti-holomorphic'' parts for the space generated by the $m^i$ and
$\bar{m}^i$ is mantained under parallel transport by $k^a$.

The solution of \eqref{eq:intgeneral} can also be obtained from the
method of characteristics. The result is an implicit function of $D-1$ 
complex arguments:
\be
F\left( \{Y^i\};\{vY^i-z_i\};u+\sum_i\bar{z}_iY^i\right)=0
\ee
one sees that the number of arguments is much less than the number of 
components of pure spinors in $D$ dimensions
\cite{Hughston:1988nz}. This should be seen as a function of the
Grassmann plane $Gr(2,D-2)$ in spinorial space, so the generic function of
the pure spinor has to be modded out by the subset of the isometries of
$SO(D-1,1)$ which leaves $Gr(2,D-2)$ invariant. These spaces are not
projective, so the twistor analysis has to be carried over using other
methods. This shows that real space applications of twistor methods in
higher dimensions are not trivial, the generic analysis based on
(complex) pure spinors is incomplete.

\section{Six Dimensional Twistors}

As seen above, an isotropic 3-plane is represented by a simple 3-form  
$T_{abc}=T_{[abc]}=e^1\wedge e^2 \wedge e^3$, whose translation to the
spinor language is the product $\kappa^A\kappa^B$, with $e^i\equiv
\kappa^{[A}(\kappa_i)^{B]}$. Frobenius' theorem states that this plane
will be integrable if the 3-form $T_{abc}$ is, up to scale, a harmonic
(closed and co-closed)\footnote{\footnotesize In this guise the theorem
is sometimes called Mariot-Robinson's.}. Since an isotropic 3-plane
is necessarily self-dual, then one needs only to check for closedness. In
terms of spinors, this condition is 
\be
\nabla_{AB}(\kappa^A\kappa^C)=0,
\ee 
or 
\be
\kappa^A\nabla_{AB}\kappa^C=-(\nabla_{AB}\kappa^A)\kappa^C
\label{eq:intcond}.
\ee
The six-dimensional twistor equation 
\be
\nabla_{AB}\kappa^C=\pi_{[A}\delta^C_{B]}\label{eq:twistoreq}
\ee
solves the constraint if $\pi_A\kappa^A=0$. The twistor equation is
the only plausible simplification for the derivative of a spinor. In
general, the derivative of $\kappa^A$ will have the following
expansion: 
\be
\nabla_{AB}\kappa^C=\pi_{[A}\delta^C_{B]}+\Theta_{AB}^C,
\ee
with $\Theta_{AB}^C$ in the ${\bf 20}$ of $SU(4)$, being antisymmetric
in the lower indices and traceless. Imposing that $\Theta_{AB}^C=0$ is 
the only Lorentz-invariant condition that leaves the right number of
degrees of freedom involved in the Kerr theorem.  

The six-dimensional twistor equation \eqref{eq:twistoreq} has the usual 
integrability condition stemming from the Ricci identity: 
\be
(\nabla_{AB}\nabla_{CD}-\nabla_{CD}\nabla_{AB})\kappa^E= {{\cal
R}_{ABCDF}}^E\kappa^F.
\ee
Contracting with the antisymmetric tensor, one obtains
\be
\epsilon^{BCDG}(\nabla_{AB}\nabla_{CD}-\nabla_{CD}\nabla_{AB})\kappa^E
= {R_{AF}}^{GE}\kappa^F,
\ee
which, using \eqref{eq:twistoreq} and rearranging the indices, can be
written as
\be
3\epsilon^{CBEG}\nabla_{[AB}\pi_{C]}+
2\epsilon^{BCEG}\nabla_{BC}\pi_A- 
\epsilon^{CDBG} \delta^E_A \nabla_{CD}\pi_B= 
2{R_{AF}}^{GE}\kappa^F.
\label{eq:twistorintegra}
\ee
Contracting $A$ and $E$, and using that
${R_{AF}}^{GA}=-15\Lambda\delta_F^G$ \eqref{eq:riemanndecomp}, we find 
that: 
\be
\epsilon^{ABCD}\nabla_{AB}\pi_C=6\Lambda\kappa^D.
\ee
with $\Lambda$ related to the scalar curvature. For Einstein manifolds
${\Phi_{AF}}^{EF}=0$, $\Lambda$ is a constant and the antisymmetric
part of the left hand side imposes a dual twistor equation for $\pi_A$:  
\be
\nabla^{AB}\pi_C=2\Lambda \delta_C^{[A}\kappa^{B]}. 
\label{eq:dualtwistor}
\ee
The twistor pair $(\kappa^A,\pi_B)$ then solves the zero mass
half-spin (Weyl) equations. The symmetric part of $EG$ in
\eqref{eq:twistorintegra} gives an algebraic condition on the Weyl
spinor: 
\be
{\Psi_{AF}}^{EG}\kappa^F=0.
\ee
Because of \eqref{eq:dualtwistor}, $\pi_c$ obeys a simmilar condition:
${\Psi_{AF}}^{EG}\pi_G=0$. In quaternionic language, the twistor
$(\kappa^A,\pi_C)$ is an eigenvector of the (symmetric) Weyl operator
with zero eigenvalue.

If the manifold has a vanishing Riemann tensor, then $\pi_C$ is
covariantly constant and therefore the twistor equation 
\eqref{eq:twistoreq} can be solved as: 
\be
\kappa^A=\psi^A+x^{AB}\lambda_B
\ee
with $\psi^A$ and $\lambda_B$ constant spinors. This is an exact
analogue of the four dimensional case. Like its 4-dimensional
sibling, it can be defined as the null subspace of the twistor space,
where the metric is the usual pairing:  
\be
\langle (\pi^A,\lambda_B), (\chi^C,\zeta_D) \rangle =
\pi^A\zeta_A+\lambda_B\chi^B.
\ee
With the choice of charge conjugation operator defining a real form in
the eight-dimensional complex space spanned by the twistors.  

For generic Einstein manifolds, the pair $Z^{\cal I}=(\psi^A,\chi_B)$
has the interpretation of a Killing spinor
\cite{vanNieuwenhuizen:1983wu}. The twistor equation and the
corresponding dual \eqref{eq:dualtwistor} are written with the help of
gamma matrices as:  
\be
\Dslash \eta = cR\eta.
\label{eq:killingspinor}
\ee
For some constant $c$, and $R$ being the scalar curvature. One then
sees that, as in four dimensions, the algebraic property that
$\kappa^A$ is annihilated by the Weyl spinor -- a principal spinor --
is imperative for the existence of solutions to
\eqref{eq:killingspinor}.

Given that one can associate spinors to integrable harmonic 3-planes,
the number of solutions to the twistor equation is a topological
invariant, the intersection number (Betti number for half-dimensional
submanifolds). Being harmonic, the forms can be seen to saturate the
Green's inequality. This property can be generalized to ``impure
spinors'', which are related to lower dimensional isotropic manifolds,
and be used to define ``calibrations'' \cite{Harvey1990} and are at
the heart of the saturation of inequalities that allow for
computations in the so-called ``attractor mechanism'' \cite{Sen:2007qy}.

\subsection{A family of Examples}

There is a family of spaces allowing for a principal spinor. These are
the Kerr-Schild spaces, whose metric is given by: 
\be
g_{ab}=\bar{g}_{ab}-2Sk_ak_b
\ee 
where $\bar{g}_{ab}$ is a maximally symmetric metric (for $dS_6$,
$AdS_6$ or $\mathbb{R}^{(5,1)}$), $k_a$ is a geodesic, shear-free null
vector field, with respect to $\bar{g}_{ab}$ and $S$ is a
function. These metrics have been studied extensively in four
dimensions \cite{Gurses1974,Stephani2003b,Ortaggio:2008iq}. The treatment
below will follow \cite{Batista:2012cp} closely. Solutions of Einstein's
equations of this form in four dimensions include all pp-wave backgrounds,
as well as the Kerr-Newman family of black holes. The properties that
$k^a$ is geodesic and shear free in the metric $\bar{g}_{ab}$ translate to
the metric $g_{ab}$. We will assume that $DS=k^c\nabla_cS=0$. If
$C^c_{ab}$ is the relative connection between the derivatives $\nabla_a$
and $\bar{\nabla}_a$, associated respectively to the metrics $g_{ab}$ and
$\bar{g}_{ab}$, we have
\be
C^c_{ab}=(\bar{\nabla}_aS)k^dk_b+(\bar{\nabla}_bS)k^dk_a-(\bar{\nabla}
^dS)k_ak_b+2S(DS)k^ck_ak_b.
\ee
Indices are raised with the maximally symmetric metric $\bar{g}_{ab}$. 
These satisfy $C^c_{ab}k^b=0$, so if $k^a$ is geodesic with respect
to $\nabla_a$, it will also be so with respect to
$\bar{\nabla}_a$. Moreover, since $\bar{g}_{ab}$ is maximally
symmetric, it is conformally flat and the integrability properties of
$k^a$ which make for the hypothesis of the Kerr theorem remain valid,
at least locally. The $k^a$ can be thought of as the real part of an
integrable 3-plane, as in Section \ref{se:kerr}. 

With a little effort, one can compute the Riemann tensor associated
with $g_{ab}$: 
\be
{R_{abc}}^d ={\bar{R}_{abc}}{}^d-\bar{\nabla}_aC^d_{bc}+\bar{\nabla}_b
C^d_{ac}+ C^e_{ac}C^d_{eb}-C^e_{bc}C^d_{ea},
\ee
with $\bar{R}_{abc}{}^d$ maximally symmetric. The result is:
\be
R_{abcd} = \bar{R}_{abcd}
-{\textstyle
\frac{R}{15}}S(k_{[a}\bar{g}_{b]c}k_d-k_{[a}\bar{g}_{b]d}k_c)-
2k_{[a}(\bar{\nabla}_{b]}\nabla_cS)k_d-2k_{[a}(\bar{\nabla}_{b]}
\nabla_dS)k_c.
\ee
The Ricci tensor is:
\be
R_{ab}=\bar{R}_{ab}-\Delta S k_ak_b
\ee
where $\Delta$ is the Laplacian in the transverse space, related to 
$\bar{g}_{ab}$. We will assume that the space is Einsteinian. Given the
form of the Riemann tensor, one can show that the Weyl tensor has the
property: 
\be
C_{abcd}k^bk^d=0,
\label{eq:triplerepeated}
\ee
which can be translated to the spinorial language:
\be
{\Psi_{AE}}^{IJ}=
C_{[AB]\,[CD]\,[EF]\,[GH]}\epsilon^{BCDI}\epsilon^{FGHJ}. 
\ee
Now, we write $k^a=\kappa^{[A}\bar{\kappa}^{B]}$, for an integrable
$\kappa^A$ satisfying \eqref{eq:twistoreq}. We choose an spinorial
basis such that
$\epsilon^{ABCD}=\kappa^{[A}\bar{\kappa}^B(\pi^*)^C(\bar{\pi}^*)^{D]}$
as in Section \ref{se:6dspinors}. Here the asterisk denotes dual. Given
that the dual spinor $\pi_B$ satisfies a similar equation
\eqref{eq:dualtwistor}, we will also assume that it is integrable. We
introduce the dual basis $\{(\kappa^*)_A, (\bar{\kappa}^*)_A, \pi_A,
\bar{\pi}_A\}$, and note that $k_a= \pi_{[A}\bar{\pi}_{B]}$. We find that
\eqref{eq:triplerepeated} means: 
\be
{\Psi_{AE}}^{IJ}\pi_{[I}\bar{\pi}_{B]}\pi_{[J}\bar{\pi}_{F]}= 0. 
\ee
A similar calculation with the $\pi_E$ results in
\be
{\Psi_{AE}}^{IJ}\kappa^{[A}\bar{\kappa}^{H]}\kappa^{[E}\bar{\kappa}^{K]}=
0. 
\ee
These identities constraint the form of the Weyl spinor so that
${\Psi_{AE}}^{IJ}\kappa^E=0$. The differential equation for $S$ can be
solved by usual methods, and the $\kappa^A$ are taken from the
general solution in flat space \eqref{eq:solutionkerr}.

As in four dimensions, this
family of spacetimes can cover all known families of reduced holonomy
manifolds. Some of those, like the cone of Einstein-Sasaki's
\cite{Martelli:2006yb}, can be obtained from Wick rotation of
Lorentzian manifolds. However, there seems to be more parameters
encoded in the $\kappa^A$ as in \eqref{eq:projtwistors}, and the
constants fixing $S$ than
the usual $p,q,r$ involved in the $L^{(p,q,r)}$ and the Reeb
vector. Although it should be remembered that in the Euclidian case
one would like to impose that the orbits of the vector $k^a$ are
closed. We hope to address this issue in the future. 

\section{Conclusions}

In this letter we addressed the relationship between Kerr theorem and
pure spinors in six dimensions. We showed that, while the spinorial
language is natural to talk about solutions to geodesic, shear-free,
null vector fields, the correspondence is not one-to-one, which is
reminiscent of the quaternionic structure of supersymmetry in six
dimensions \cite{Kugo:1982bn}. We then turned to the problem of the
twistor equation in six dimensions, and showed that, just like the
four dimensional case \cite{vanNieuwenhuizen:1983wu}, the Killing
spinor equation can be cast in the Newman-Penrose language quite
naturally. Finally, we presented a generalization of Kerr-Schild
metrics which display the algebraic property allowing for solutions of
the twistor equation.  

A great deal of applications of pure spinors has been put forward in
the last ten years, mainly to the covariant superstring
\cite{Berkovits:2000fe}. The effort has been basically centered on
algebraic aspects of pure spinors. However, most aspects of such
applications have had their birth in geometrical aspects of four
dimensional general relativity. This work is the result of some effort
spent in trying to understand pure spinors from a geometrical point of
view in six dimensions. In eight or more dimensions, the geometrical
point of view becomes more natural, as the algebraic one becomes more
mysterious: for instance, the pure condition becomes a quadratic
constraint on spinors, while the relationship to maximal dimension
isotropic planes continue to hold. In six dimensions, as we can see
here, many of the features translate quite naturally to the new
setting. We hope that the fresh point of view will be useful to other
people. 

\section*{Acknowledgements}

The author is indebted to Carlos Batista and Amílcar de Queiroz for a
critical and thorough, reading of the manuscript, and would also like
to thank A. P. Balachandran and Fábio Novaes for comments and
suggestions. 

\providecommand{\href}[2]{#2}\begingroup\raggedright
\endgroup
\end{document}